\documentclass{article}

\title{Some Remarks on Neutrino Oscillation and Masses}

\author{A. Zee\thanks{zee@kitp.ucsb.edu}\\
Kavli Institute for Theoretical Physics \\
University of California\\
Santa Barbara, CA 93106, USA}

\begin{document}

\maketitle

\begin{abstract}
We discuss an Ansatz for the neutrino mixing matrix and speculate on the
form and origin of the neutrino mass matrix.
\end{abstract}

\section{Mixing Matrix}

Thanks to heroic experimental efforts, the neutrino mixing matrix
has now been determined to be\cite{expt} (see the various experimental talks
at this conference)
\begin{equation}
V=\left(
\begin{array}{lll}
0.72-0.88 & 0.46-0.68 & <0.22 \\
0.25-0.65 & 0.27-0.73 & 0.55-0.84 \\
0.10-0.57 & 0.41-0.80 & 0.52-0.83
\end{array}
\right)  \label{exp}
\end{equation}
(The numbers given are absolute values of the matrix elements of V.)
The mixing matrix $V$ relates the neutrinos current eigenstates (denoted by $%
\nu _{\alpha }$ ($\alpha =e,$ $\mu ,$ $\tau )$ and coupled by the $W$ bosons
to the corresponding charged leptons) to the neutrino mass eigenstates
(denoted by $\nu _{i}$ ($i=1,2,3))$ according to
\begin{equation}
\left(
\begin{array}{l}
\nu _{e} \\
\nu _{\mu } \\
\nu _{\tau }
\end{array}
\right) =V\left(
\begin{array}{l}
\nu _{1} \\
\nu _{2} \\
\nu _{3}
\end{array}
\right)
\end{equation}

We will take the neutrinos to be Majorana\cite{kay} as seems likely, so that
we have in the Lagrangian the mass term
\begin{equation}
\mathcal{L}=-\nu _{\alpha }M_{\alpha \beta }C\nu _{\beta }+h.c.
\end{equation}
where $C$ denotes the charge conjugation matrix. Thus, the neutrino mass
matrix $M$ is symmetric. For the sake of simplicity we will assume $CP$
conservation so that $M$ is real. Indeed, we have already written all the
entries as real in (\ref{exp}).

With this simplification, $M$ is diagonalized by an orthogonal
transformation
\begin{equation}
V^{T}MV=\left(
\begin{array}{lll}
m_{1} & 0 & 0 \\
0 & m_{2} & 0 \\
0 & 0 & m_{3}
\end{array}
\right)  \label{diag}
\end{equation}

Clearly, we are free to multiply $V$ on the right by some diagonal matrix
whose diagonal entries are equal to $\pm 1$. This merely multiplies each of
the columns in $V$ by an arbitrary sign. Various possible phases have been
discussed in detail in the literature.\cite{nieves}

We could suppose either that the entries in $V$ represent a bunch of
meaningless numbers or that they point to some deeper structure or symmetry.
In the latter spirit, let us make a guess of what $V$ might be.

Since $V_{e3}$ appears to be small, let us boldly set it to $0$. Next, since
$1/\sqrt{2}\sim 0.707$ we will guess that $V_{\mu 3}=(0.55-0.84)=1/\sqrt{2}.$
Finally, since $1/\sqrt{3}\sim 0.577$ we will set $V_{e2}=(0.46-0.68)=1/%
\sqrt{3}.$ In other words, we propose that we know the upper triangular
entries of the matrix $V:$

\[
V=\left(
\begin{array}{rrr}
{X} & {\frac{1}{\sqrt{3}}} & 0 \\
{X} & {X} & {\frac{1}{\sqrt{2}}} \\
{X} & {X} & {X}
\end{array}
\right),
\]
where ${X}$ denotes an unknown quantity.

Remarkably, this essentially fixes the mixing matrix $V$. Once we take the
last column to be proportional to $(0,1,-1)$, orthogonality and our
``knowledge'' that $V_{e2}$ is $1/\sqrt{3}$ immediately fix the second
column to be proportional $(1,1,1)$ and hence the first column to be
proportional to $(-2,1,1)$. X. G. He and I therefore arrived at the Ansatz
or guess\cite{hz1}
\begin{equation}
V=\left(
\begin{array}{rrr}
-{\frac{2}{\sqrt{6}}} & {\frac{1}{\sqrt{3}}} & 0 \\
{\frac{1}{\sqrt{6}}} & {\frac{1}{\sqrt{3}}} & {\frac{1}{\sqrt{2}}} \\
{\frac{1}{\sqrt{6}}} & {\frac{1}{\sqrt{3}}} & {-\frac{1}{\sqrt{2}}}
\end{array}
\right) .  \label{theV}
\end{equation}

This mixing matrix (but curiously, with the first and second column
interchanged) was first suggested by Wolfenstein more than 20 years
ago\cite{20}. Later it was proposed by Harrison, Perkins and
Scott\cite{21}, and
subsequently studied extensively by them and by Xing\cite{22}.

The mixing matrix $V$ may be factorized as $V=V_{23}V_{12}$ where
\begin{equation}
V_{23}=\left(
\begin{array}{rrr}
1 & {0} & 0 \\
{0} & {\frac{1}{\sqrt{2}}} & {\frac{1}{\sqrt{2}}} \\
{0} & {\frac{1}{\sqrt{2}}} & {-\frac{1}{\sqrt{2}}}
\end{array}
\right)
\end{equation}
and
\begin{equation}
V_{12}=\left(
\begin{array}{rrr}
-\sqrt{\frac{2}{3}} & {\frac{1}{\sqrt{3}}} & 0 \\
{\frac{1}{\sqrt{3}}} & \sqrt{\frac{2}{3}} & {0} \\
{0} & {0} & {1}
\end{array}
\right) .
\end{equation}
In other words, if we follow Wolfenstein and define $\nu _{x}\equiv (\nu
_{\mu }+\nu _{\tau })/\sqrt{2}$ and $\nu _{y}\equiv (-\nu _{\mu }+\nu _{\tau
})/\sqrt{2}$ we find that the mass eigenstates are given by
\begin{equation}
\nu _{1}=-\sqrt{\frac{2}{3}}\nu _{e}+{\frac{1}{\sqrt{3}}}\nu _{x},
\end{equation}
\begin{equation}
\nu _{2}={\frac{1}{\sqrt{3}}}\nu _{e}+\sqrt{\frac{2}{3}}\nu _{x},
\end{equation}
and
\begin{equation}
\nu _{3}=\nu _{y}
\end{equation}

I find this matrix $V$ rather attractive, but how could we obtain such an
``elegantly simple'' mixing matrix?

Recently, Harrison, Perkins and Scott\cite{hps} proposed a discrete symmetry
group $\mathcal{D}$ and managed to obtain $V.$ Unfortunately, as emphasized
by Low and Volkas\cite{lv}, they have to allow the left handed neutrinos and
the left handed charged leptons, which of course belong to the same doublet
under the standard $SU(2)\otimes U(1),$ to transform differently. Thus the
low energy symmetry group of the electroweak interaction proposed by
Harrison et al could not have the form $SU(2)\otimes U(1)\otimes \mathcal{D}%
. $ Assuming that it indeed has this form and assuming only one Higgs
doublet, Low and Volkas went on and proved a no-go theorem showing that no
choice of $\mathcal{D}$ would lead to $V.\bigskip $

\section{Mass Matrix}

Neutrino oscillation\cite{osc} experiments can only determine the absolute
value of the mass squared differences $\Delta m_{ij}^{2}\equiv $ $%
m_{i}^{2}-m_{j}^{2}.$ At the 99.3\% confidence level $\Delta m_{ij}^{2}$ are
determined by
\[
1.5\times 10^{-3}eV^{2}\leq |\Delta m_{32}^{2}|\leq 5.0\times 10^{-3}eV^{2},
\]
and
\[
2.2\times 10^{-5}eV^{2}\leq |\Delta m_{21}^{2}|\leq 2.0\times 10^{-4}eV^{2},
\]
with the best fit values given by
\begin{equation}
|\Delta m_{32}^{2}|=3.0\times 10^{-3}eV^{2}  \label{32}
\end{equation}
and
\begin{equation}
|\Delta m_{21}^{2}|=7.0\times 10^{-5}eV^{2}.  \label{21}
\end{equation}
Thus, we could have either the so-called normal hierarchy in which $%
|m_{3}|>|m_{2}|\sim |m_{1}|$ or the inverted hierarchy $|m_{3}|<|m_{2}|\sim
|m_{1}|.$ At present, we have no understanding of the neutrino masses just
as we have no understanding of the charged lepton and quark masses.

In general, when we diagonalize a matrix $M$ as in (\ref{diag}) we expect
the eigenvalues $m_{i}$ and the matrix $V$ to depend on the matrix elements $%
M_{\alpha \beta }.$ Only a certain class of matrices would have the property
(which Low and Volkas called ``form-diagonalizable'') such that $V$ comes
out as a matrix of pure numbers as in (\ref{theV}). At first sight, it seems
a bit odd that form-diagonalizable matrices exist, but a moment's thought
indicates that they could be constructed as follows: given three orthonormal
column vectors $\vec{v}^{(i)}=\{v_{\alpha }^{(i)}\}$ whose components are
pure numbers, then $M=\sum_{i=1}^{3}m_{i}$ $\vec{v}^{(i)}(\vec{v}^{(i)})^{T}$
is form diagonalizable for arbitrary $m_{i}.$ Thus, if we believe in (\ref
{theV}) then the neutrino mass matrix is given by
\begin{equation}
M=\frac{m_{1}}{6}\left(
\begin{array}{lll}
4 & -2 & -2 \\
-2 & 1 & 1 \\
-2 & 1 & 1
\end{array}
\right) +\frac{m_{2}}{3}\left(
\begin{array}{lll}
1 & 1 & 1 \\
1 & 1 & 1 \\
1 & 1 & 1
\end{array}
\right) +\frac{m_{3}}{2}\left(
\begin{array}{lll}
0 & 0 & 0 \\
0 & 1 & -1 \\
0 & -1 & 1
\end{array}
\right)
\end{equation}

The three column vectors contained in $V$ are the eigenvectors of the matrix
\begin{equation}
M_{0}=a\left(
\begin{array}{rrr}
2 & {0} & 0 \\
{0} & {-1} & {3} \\
{0} & {3} & -1
\end{array}
\right) ,
\end{equation}
with eigenvalues $m_{1}=m_{2}=2a,$ and $m_{3}=-4a.$ (The parameter $a$
merely sets the overall scale.) Thus, $\Delta m_{21}^{2}=0$ and this pattern
reproduces the data $|\Delta m_{21}^{2}|/|\Delta m_{32}^{2}|\ll 1$ to first
approximation. Because of the degeneracy in the eigenvalue spectrum, $V$ is
not uniquely determined. We can always replace $V$ by $VW$ where

\[
W=\left(
\begin{array}{ll}
R & 0 \\
0 & 1
\end{array}
\right),
\]
with $R$ a $2 \times 2$ rotation matrix. To determine $V,$ and at the same
time to split the degeneracy between $m_{1}\ $and $m_{2},$ we perturb $M_{0}$
to $M=M_{0}+\delta M_T,$ where

\[
\delta M_{T}=\varepsilon a\left(
\begin{array}{lll}
0 & 1 & 1 \\
1 & 0 & 1 \\
1 & 1 & 0
\end{array}
\right) .
\]

We have the mass eigenvalues $m_{1}=2a(1-\varepsilon/2
),m_{2}=2a(1+\varepsilon ),$ and $m_{3}=-4a(1+\varepsilon/4).$ Thus, we can
determine to the lowest order $\varepsilon = \Delta m_{21}^{2}/\Delta
m_{32}^{2}$. The overall scale of the mass matrix $a$ is given by $a^2 =
\Delta m^2_{32}/12$.

Other perturbations can also lead to the same mixing matrix $V$ while
splitting the degeneracy $\Delta m_{21}^{2}=0$. An interesting example is
the `` democratic'' form

\[
\delta M_{D}=\varepsilon a\left(
\begin{array}{lll}
1 & 1 & 1 \\
1 & 1 & 1 \\
1 & 1 & 1
\end{array}
\right) .
\]
The matrix $\delta M_{D}$ is evidently a projection matrix that projects the
first and third columns in $V$ to zero. Thus, the eigenvalues are given by $%
m_{1}=2a,m_{2}=2a(1+3\varepsilon /2),$ and $m_{3}=-4a,$ where to the lowest
order $\varepsilon =\Delta m_{21}^{2}/\Delta m_{32}^{2}$ and $a^{2}=\Delta
m_{32}^{2}/12$. We note that this mass matrix is not traceless.

We mention that there is a whole class of models we can propose. Generalize $%
M_0$ to

\[
\tilde{M}_{0}=a\left(
\begin{array}{ccc}
2 & 0 & 0 \\
0 & 1-y & 1+y \\
0 & 1+y & 1-y
\end{array}
\right) ,
\]
with the case mentioned earlier corresponding to $y=2$. Thus in general we
propose

\[
M=\tilde M_0+ \delta M,
\]
with $\delta M$ being $\delta M_T$ or $\delta M_D$. They lead to the same
mixing matrix $V$, with the eigenvalues $m_i$ given by $(2a(1-\varepsilon
/2), 2a(1+\varepsilon), -2a(y +\varepsilon/2))$ and $(2a, 2a(1+3\varepsilon
/2), -2ay)$, respectively.

Note that the most general mass matrix which produces the mixing matrix V
can be expressed as linear combinations of the three matrices of the forms
given by $M_{0}$, $\delta M_{T}$ and $\delta M_{D}$. Once we committed to a
specific form for $M$, the three parameters specifying the linear
combination merely parametrize the three neutrino masses $m_{1,2,3}$.
Obviously for any given mixing matrix, the mass matrix can be specified by
mass eigenvalues.

\bigskip With 2 experimental numbers for the 3 masses $m_{1,2,3}$, we have
to make another wild guess in order to determine $M.$ Since $%
M=M_{0}+\delta M_{T}$ gives a reasonable fit to the data and since it is
traceless, one may be tempted to conjecture that this property provides a
clue to the origin of the neutrino mass matrix. In a recent paper\cite{hz2},
X. G. He and I gave a phenomenological analysis of the data imposing the
condition\cite{bfns} $TrM=0$ without speculating
on its theoretical origin.

If there is no $CP$ violation in the neutrino mass matrix $M$, the mass
matrix can always be made real and it can be diagonalized by an orthogonal
transformation. In this case the traceless condition $TrM=0$ is equivalent
to the ``zero sum'' condition
\begin{equation}
m_{1}+m_{2}+m_{3}=0.
\end{equation}
(But if $CP$ is not conserved, the ``zero sum'' and traceless conditions are
different. One needs to be careful about the phase
definitions\cite{nieves}.)
We note that the traceless condition holds if $M=[A,B]$, that is, the
mass matrix can be expressed as a commutator of two matrices $A$ and $B$.

As remarked earlier, we are free to choose the signs of the column vectors
in the mixing matrix and to make chiral rotations on the neutrino fields to
change the relative signs of the mass eigenvalues. Without information on
the relative signs of the eigen-masses, the column vectors can only be
determined up to $\pm i$. This can be expressed by multiplying a diagonal
phase matrix $P=Diag(e^{i\sigma },e^{i\rho },1)$ to the right of $V$. With $%
CP$ invariance, $\sigma $ and $\rho $ can take the values of zero or $\pm
\pi /2$. Neutrinoless double beta decays will provide some crucial
information on these phases.

Combining the zero mass condition with the experimental data on the
differences of mass squared we find that the mass eigenvalues exhibit
two types of hierarchies,
\begin{eqnarray}
&&(I)\;\;\;m_{3}\approx -2m_{1}\approx -2m_{2}\approx 0.064\;eV  \nonumber \\
&&(II)\;\;m_{1}\approx -m_{2}\approx 0.054\;\;eV,\;\;\mbox{and}%
\;\;m_{3}\approx 0.00064\;\;eV.  \label{hierarchy}
\end{eqnarray}
The sign of $\Delta m_{32}^{2}$ decides which mass hierarchy the solutions
belong to. Note that the ``natural'' sign $\Delta m_{32}^{2}>0$ corresponds
to scenario $(I)$, in which the masses are of the same order of magnitude,
in contrast to scenario $(II)$, in which $m_{3}$ is two order of magnitude
smaller than $m_{1}$ and $m_{2}$. We would like to suggest that $(I)$ is
favored over $(II)$.

Our purpose here is evidently not to give a detailed fit to the data, but to
suggest some relatively simple and appealing mass matrices. The appearance
of simple integers in the mixing and mass matrices we proposed is perhaps
intriguing and provides a glimmer of a hope that they may be obtained by
group theoretic considerations. To provide a theoretical origin of the mass
matrix $M$ presents an interesting challenge.\medskip

\section{Theory}

Theoretically, pitifully little is known about neutrino masses and mixing.
The only firm theoretical statement is that since the standard
model is correct at low energies, neutrino masses have to come from the
following dimension 5 operator in the
Lagrangian,\cite{wein, wz, weldonz}
\begin{equation}
\mathcal{L}=\frac{1}{\mathcal{M}}(\varepsilon _{ij}\psi _{i}\varphi
_{j})C(\varepsilon _{kl}\psi _{k}^{^{\prime }}\varphi _{l}^{\prime }).
\end{equation}
Here $\psi \ $and $\psi ^{\prime }$ denote left handed lepton doublets and $%
\varphi $ and $\varphi ^{\prime }$ Higgs doublets, and $i,j,k,l$ denote $%
SU(2)$ indices. The unknown mass parameter $\mathcal{M}$ sets the mass scale
of the new physics responsible for generating the neutrino masses.

Life is full of bifurcating choices: so too the neutrino mass model builder
is immediately faced with the choice of introducing right handed neutrinos
or not. A neutrino could have either a Dirac mass or a Majorana mass. In the
first alternative, one needs to introduce right handed neutrino fields and
the question immediately arises on why the neutrino Dirac masses are so
small compared to the charged lepton masses in the theory. This question was
answered elegantly by the see-saw mechanism, in which the right handed
neutrino fields are given large Majorana masses. But if we are willing to
introduce Majorana masses for the right handed neutrino fields, perhaps we
should consider dispensing with right handed neutrino fields altogether and
simply try to generate Majorana masses for the existing left handed neutrino
fields. We shall try to generate this Majorana masses through quantum
mechanical effect. This has the added advantage of having naturally small
neutrino mass and the new physics at a potentially experimentally accessible
scale.

Since in the standard model, the left handed neutrino fields belong to
doublets $\psi _{aL}$ (with $a$ a family index) we cannot simply put in
Majorana mass terms. The general philosophy
followed in Ref.\cite{zee1,zee2,zee3} is that we should
feel freer to alter the scalar field
sector than other sectors since the scalar field sector is the least
established one in the standard model. Out of the doublets we can form the
Lorentz scalar $(\psi _{aL}^{i}C\psi _{bL}^{j})$ (where $i,j$ denote
electroweak $SU(2)$ indices and $C$ the charge conjugation matrix): this can
be either a triplet or a singlet under $SU(2)$. If we couple a triplet field
to this lepton bilinear, then when the neutral component of the triplet
field acquires a vacuum expectation value, the neutrinos immediately acquire
Majorana masses. We considered this model unattractive: not only does it
lack predictive power, but the rather accurately studied ratio of $W$ and $Z$
boson masses puts a stringent bound on any triplet Higgs. In addition, there
is no natural way to explain the smallness required of this vacuum
expectation value.

We thus chose the alternative of coupling to an $SU(2)$ singlet (charged)
field $h^{+}$ via the term $f^{ab}(\psi _{aL}^{i}C\psi _{bL}^{j})\varepsilon
_{ij}h^{+}.$ An interesting point is that due to Fermi statistics the
coupling $f^{ab}$ must be anti-symmetric in $a$ and $b.$ We are forced to
couple leptons in one family to leptons in another one. Thus, the term above
contains $f^{e\mu }(\nu _{e}C\mu ^{-}-e^{-}C\nu _{\mu })h^{+},$ for instance$%
.$ The term $f^{ab}(\psi _{aL}^{i}C\psi _{bL}^{j})\varepsilon _{ij}h^{+}$ in
itself does not violate lepton number $L$ since we can always assign $L=-2$
to $h^{+}.$ But we note that we can also couple $h^{+}$ to the Higgs
doublets via $M_{\alpha \beta }\phi _{\alpha }\phi _{\beta }h^{+}$ if there
are more than one Higgs doublet. By Bose statistics, the coupling matrix $%
M_{\alpha \beta }$ is antisymmetric and thus we need to have at least two
Higgs doublets. We do not regard the necessity of more than one Higgs 
doublet as
unattractive. Indeed, theorists have always been motivated by one reason or
another to introduce additional Higgs doublets, and surely in the debris of
breaking down from some high mass scale physics there would be numerous
scalar fields. If the doublets are required to have zero lepton number by
their respective Yukawa couplings, the term
$M_{\alpha \beta }\phi _{\alpha }\phi _{\beta }h^{+}$ violates lepton number
by two
units, just right for generating neutrino Majorana masses.

{From} general principles we know that neutrino Majorana masses must be
generated and that they must come out as finite, that is, calculable in
terms of the parameters of the theory. Indeed, it is easy to see that
calculable neutrino Majorana masses are generated by quantum fluctuations in
one loop\cite{zee1}.

There has been a considerable literature\cite{13} devoted to studying the
various implications of this model, known as the Zee model. Soon after this
model was proposed, Wolfenstein\cite{wolf} suggested an interesting
simplification by imposing a discrete symmetry so that one of the two
minimally necessary Higgs doublets does not couple to leptons. This
simplified Zee model, or the Zee-Wolfenstein model,
is now ruled out\cite{framp}.
The original Zee model, however, continues to be phenomenologically
viable\cite{hase, henew}. There are also several possible
variants\cite{changz} of this model.
It would be interesting to see if the mixing and
mass matrices discussed in the first half of this paper could possibly
emerge from this class of models.

\section*{Acknowledgments}

Part of the material discussed in this paper is based on work done with X.
G. He. I thank R. Volkas, C. N. Leung, and S. Tovey for inviting me to speak
at this conference. I would also like to thank J. Nieves for computer 
help. This
work was supported in part by the National Science Foundation under grant
number PHY 99-07949.

\end{document}